# PageRank Algorithm using Eigenvector Centrality- New Approach

Saumya Chandrashekhar Suvarna
Computer Science Department
Vellore Institute of Technology
suvarna.saumyacjyoti2014@vit.ac.in

Mashrin Srivastava
Computer Science Department
Vellore Institute of Technology
mashrin.msrivastava2014@vit.ac.in

Prof. B Jaganathan
Mathematics Department
Vellore Institute of Technology
jaganathan.b@vit.ac.in

Dr. Pankaj Shukla
Mathematics Department
Vellore Institute of Technology
pankaj.shukla@vit.ac.in

*Abstract*—**The purpose of the research is to find a centrality measure that can be used in place of PageRank and to find out the conditions where we can use it in place of PageRank. After analysis and comparison of graphs with a large number of nodes using Spearman's Rank Coefficient Correlation, the conclusion is evident that Eigenvector can be safely used in place of PageRank in directed networks to improve the performance in terms of the time complexity.**

## I. INTRODUCTION

In today's era of computer technology with the vast usage of the World Wide Web, users want a fast and accurate answer. It is the need of the hour to explore every avenue for quicker results and better time complexity. Thus a constant re-evaluation of existing algorithms is necessary. Centrality measures in any kind of network highlights the important parts of that network. In dense networks, there is a potential for more importance as compared to sparse networks. Ranking is essential in terms of finding the most important or influential nodes in a network.

## II. EXISTING ALGORITHMS[1]

Degree centrality selects the most important node based on the principle that an important node is involved in a lot of interactions. However degree centrality doesn't take into account the importance of the nodes it is connected to. Closeness centrality decides the importance of the nodes according to the concept that important nodes will be able to communicate quickly with the nodes. Betweenness centrality finds the most important node based on the theory that if a node is important it will lie on the path between two nodes. Betweenness and closeness centrality is mostly redundant when it comes to ranking of webpages as it attaches no proper importance to the relevant webpages. Degree centrality is relevant however it doesn't take into account the importance of the node pointing to it hence it gives ample opportunity for malicious and inauthentic websites to be ranked first. Apart from the abovementioned algorithms, other algorithms to find the ranking of a webpage include

### A. The Existing Algorithms in Use

Eigentrust Algorithm[2] : Peer-to-peer network consists of partitioning tasks between various peers such that each peer is equally privileged. This method is popular for sharing information but due to its open source nature, the spread of inauthentic files is easy. Eigentrust Algorithm assigns a rank to each peer based on the past uploads made by the peer in an attempt to reduce the rank of inauthentic or malicious peers trying to pose as peers distributing authentic files.

SimRank: There are many areas where the similarity of objects or interests come into play. The most obvious one being the use to find similar documents on the World Wide Web. Another is to group people with similar interests together, for example the recommended friends list by social media sites. SimRank measures the similarity between two objects based on the basis of the objects that they are referenced by.

TrustRank[3]: Web spam pages are created in order to mislead the search engines by using various techniques to get a higher rank. TrustRank involves manually identifying a set of authentic websites known as seed pages and sending a crawl to identify pages similar to the seed pages. In Anti-Trust Rank[4] unauthentic or malicious websites are found out and websites close to that are non-trustworthy websites. Trust rank is decreased as it moves father away from the seed site.

### B. Algorithms Used By GOOGLE

Google is the most popular search engine used. The reason being that it is able to provide relevant search results quickly. The very first algorithm used by Google was PageRank algorithm by Larry page and Sergey Brin. However incidents like Google bombing, where a search for a topic can lead to the webpages of a seemingly unrelated topic prompted tweaks in algorithms used by Google. Google bombs can be a done for business, political or comical reasons. It works on the principle of spamdexing, which is the manipulation of the index used by the search engines and heavy linking of websites. In order to reduce incidents like Google bombing, Google uses many algorithms along with the original PageRank algorithm to ascertain the final ranking of a webpage. These algorithms include-

Google Panda: Authority is one of the key measurements of ranking in search engines. The trust can be measured by the authority of the link to the article, where the more a link is authoritative, the more the article it points to can be trusted. Google Panda is essentially a filter of the content quality, depreciating low quality websites including websites that have unoriginal or redundant content, contains a huge number of advertisements, have content that is not written or authorized by an expert. If the panda rank score of a site is high, the website gets 'Pandified' which means that the pages in the site are imposed with a penalty.

Google Hummingbird: Google Hummingbird attempts to judge the intent of the person making the query ie. to take into consideration the meaning of the sentence as a whole as compared to certain keywords. It is said that Google uses the vast database of information or the 'Knowledge graph' to determine the best results.

Google Penguin: Google Penguin assigns a penalty to the webpages which are involved in the usage of black hat Search Engine Optimization techniques or violation of Google's webmaster guidelines. The goal is to do away with keyword stuffing, link building, meaningless and irrelevant content and doorway pages or paying for links to the website.

## III. ALGORITHMS FOR COMPARISON

### A. Existing: PageRank Algorithm

PageRank[5] centrality defines the importance of nodes based on the number and quality of nodes connected to it. The World Wide Web can be represented as a directed graph in which every webpage is represented by a node and the edges pointing to a node represents the links pointing to a webpage and the edges pointing away from the node represents the links that are pointing towards other websites. The most relevant page will be decided not only by the in-degree but also the importance of the node pointing towards the node in question. PageRank is basically based on the probability of a person surfing the web by randomly clicking on links to stop. The probability that the person will continue to click on the links is given by the damping factor and studies suggest that the ideal value is 0.85. The formula is given by

$$PR_a = (1-d) + d \sum_{b \in S_a} \frac{PR_b}{Ol_b}$$

Where $S_a$ is a set of all the webpages pointing to a page 'a' and $Ol_b$ is the out-links from the page, $d$ is the damping factor

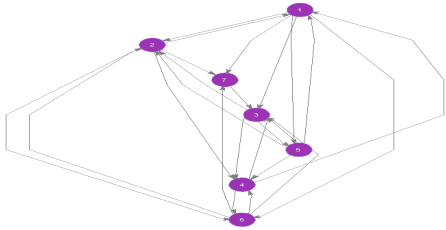

Fig. 1. A 7 node graph used to illustrate value of the centrality measures.

For the Graph given in Fig.1, the value of PageRank is given by

$$PR = \begin{bmatrix} 1.5242 \\ 1.2125 \\ 1 \\ 0.8867 \\ 1.0283 \\ 1.0283 \\ 0.3200 \end{bmatrix}$$

The time complexity of this algorithm using adjacency matrix as input is $O(V^3)$, where V is the number of vertices. The initial value of Page Rank is assumed and the subsequent values are calculated iteratively till the values converge. igraph[6] has stated that the time complexity for the algorithm with input as adjacency list is $O(|E+V|)$ where V is the total number of vertices and E is the total number of edges in the graph.

*B. Proposed: Eigenvector Centrality*

Eigenvector centrality[7] defines the importance of a node in the graph by the importance of the nodes connected to it. It can be used for finding the most influencing person or for the identification of the spread of infection. One of the most interesting applications of eigenvector centrality is its potential to be used for analyzing the connectivity patterns in the data obtained from a functional magnetic resonance scale[8].

Let us assume a graph G that has E edges and V vertices. The graph can be given by G(E,V). If A is the adjacency matrix, for the graph G, the value of $a_{ij}$ will be equal to the weight of the edges of the graph. If the graph is a directed graph, and the direction is from i to j ie. the head is at j node and the tail is at i node, then the value of $a_{ij}$ will be equal to the edge weight. If the graph is undirected, both the values of $a_{ij}$ and $a_{ji}$ will be equal to the edge weight. If two vertices are not connected, the value of $a_{ij}$ is 0.

$$Ax = \lambda x$$

The above equation represents the eigenvector centrality in the matrix form, where A is the adjacency matrix and λ is the largest eigenvalue of the adjacency matrix. x is the vector containing the centrality score of the node. The Perron-Frobenius theorem states that there is atleast one non-negative eigenvector corresponding to the largest eigenvalue of a square matrix. It can also be defined as the eigenvector corresponding to the largest eigenvalue λ of the matrix A.

$$\lambda x_k = \sum_{l=i}^{n} a_{kl} x_l \qquad k = 1,2,3\ldots\ldots n$$

The above equation represents the eigenvector centrality in a sum form where λ is the largest eigenvalue of the adjacency matrix and $a_{kl}$ represents the entry in the k$^{th}$ row and l$^{th}$ column.

For the graph given in Fig.1, the adjacency matrix will be

$$A = \begin{bmatrix} 0 & 1 & 1 & 0 & 1 & 1 & 1 \\ 1 & 0 & 0 & 1 & 0 & 1 & 1 \\ 0 & 1 & 0 & 1 & 1 & 0 & 0 \\ 1 & 0 & 1 & 0 & 0 & 1 & 0 \\ 1 & 1 & 1 & 1 & 0 & 0 & 0 \\ 0 & 1 & 1 & 1 & 0 & 0 & 1 \\ 0 & 0 & 1 & 0 & 0 & 0 & 0 \end{bmatrix}$$

The eigenvalues calculated are
$\lambda_0 = 3.3911$, $\lambda_1 = -1.1290 + 1.0560i$, $\lambda_2 = -1.1290 - 1.0560i$, $\lambda_3 = -0.0334 + 0.8151i$, $\lambda_4 = -0.0334 - 0.8151i$, $\lambda_5 = 0.1522$, $\lambda_6 = -1.2185$

The principal or the largest eigenvalue is $\lambda_0 = 3.3911$ and the corresponding eigenvector which also represents the eigenvector centrality of the graph given in Fig.1 is

$$v_0 = \begin{bmatrix} 0.4924 \\ 0.3853 \\ 0.3564 \\ 0.3548 \\ 0.4686 \\ 0.3544 \\ 0.1051 \end{bmatrix}$$

The time complexity of this algorithm using adjacency matrix as an input is $O(V^2)$, where V is the total number of vertices. The calculation of the time complexity hinges on the calculation of the eigenvector as the rest of the processes can be done in constant time. igraph has stated that the time complexity for the algorithm using adjacency list as input is $O(|V|)$. Time complexity analysis consists of the assumption that $n \to \infty$. If we consider each webpage to be a node in a graph this assumption becomes true due to the vastness of the websites that is available on the internet.

In a directed graph, in the case that every node (V) is connected to every other node (V-1), the maximum number of edges can be given by $V(V-1)$. Hence in the calculation of the time complexity of the worst case, we can take the value of E as $V^2$ hence giving the time complexity as $O(V^2)$, thus proving that the time complexity to calculate the values of the eigenvector centrality $O(V)$ is better than the existing PageRank algorithm.

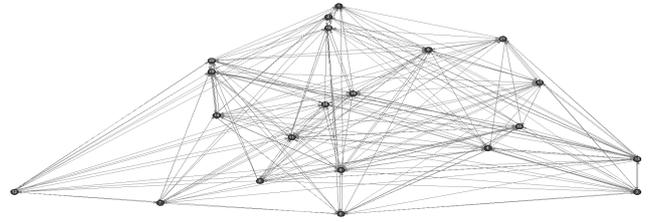

Fig. 4. A 21 node graph used for comparison of the centrality measures.

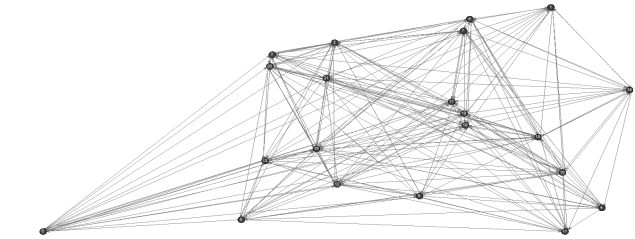

## IV. CALCULATION AND VERIFICATION

In our endeavor to prove that our hypothesis is right, we have found the eigenvector and PageRank values for several graphs. We initially started with 21 node directed graphs with each node representing a webpage. However, practically it is irrelevant to have such few webpages in a network therefore we increased the number of nodes to 50 and then 100 to simulate the trends with the increase in the number of webpages.

Fig. 3. A 21 node graph used for comparison of the centrality measures.

TABLE 1
COMPARISON BETWEEN EIGENVECTOR CENTRALITY AND PAGERANK CENTRALITY FOR GRAPH SHOWN IN FIG.3

| Vertex | Page rank Centrality | Rank | Eigenvector Centrality | Rank |
|---|---|---|---|---|
| v19 | 1.4466 | 1 | 0.3168 | 1 |
| v16 | 1.2509 | 2 | 0.273 | 3 |
| v2 | 1.2364 | 3 | 0.2818 | 2 |
| v15 | 1.1694 | 4 | 0.2517 | 5 |
| v17 | 1.1688 | 5 | 0.254 | 4 |
| v20 | 1.1065 | 6 | 0.2391 | 6 |
| v12 | 1.0696 | 7 | 0.2262 | 7 |
| v8 | 1.0356 | 8 | 0.2224 | 8 |
| v3 | 1.0296 | 9 | 0.2139 | 10 |
| v13 | 0.9872 | 10 | 0.2145 | 9 |
| v10 | 0.9800 | 11 | 0.2112 | 11 |
| v6 | 0.9744 | 12 | 0.2064 | 13 |
| v7 | 0.9407 | 13 | 0.2085 | 12 |
| v5 | 0.8944 | 14 | 0.1852 | 16 |
| v4 | 0.8915 | 15 | 0.1884 | 14 |
| v9 | 0.8843 | 16 | 0.1799 | 17 |
| v18 | 0.8658 | 17 | 0.1882 | 15 |
| v21 | 0.8261 | 18 | 0.175 | 18 |
| v14 | 0.8227 | 19 | 0.1674 | 19 |
| v1 | 0.7954 | 20 | 0.1612 | 20 |
| v11 | 0.6242 | 21 | 0.1239 | 21 |

TABLE 2
COMPARISON BETWEEN EIGENVECTOR CENTRALITY AND PAGERANK CENTRALITY FOR THE GRAPH SHOWN IN FIG. 4

| Vertex | Page rank | Rank | Eigenvector | Rank |
|---|---|---|---|---|
| v13 | 1.3884 | 1 | 0.3011 | 1 |
| v9 | 1.2609 | 2 | 0.2837 | 2 |
| v17 | 1.2511 | 3 | 0.2622 | 4 |
| v3 | 1.2085 | 4 | 0.2755 | 3 |
| v19 | 1.1892 | 5 | 0.2574 | 5 |
| v8 | 1.1586 | 6 | 0.2452 | 7 |
| v6 | 1.1387 | 7 | 0.2467 | 6 |
| v20 | 1.0904 | 8 | 0.2265 | 10 |
| v11 | 1.0795 | 9 | 0.2343 | 8 |
| v18 | 1.0750 | 10 | 0.2294 | 9 |
| v12 | 0.9454 | 11 | 0.1951 | 11 |
| v21 | 0.9312 | 12 | 0.1935 | 13 |
| v7 | 0.9183 | 13 | 0.1912 | 14 |
| Vertex | Page rank | Rank | Eigenvector | Rank |
| v10 | 0.9053 | 14 | 0.194 | 12 |
| v14 | 0.9037 | 15 | 0.1858 | 16 |
| v2 | 0.9028 | 16 | 0.1893 | 15 |
| v5 | 0.8496 | 17 | 0.1838 | 17 |
| v4 | 0.7528 | 18 | 0.1558 | 18 |
| v16 | 0.7365 | 19 | 0.1519 | 20 |
| v15 | 0.7179 | 20 | 0.1531 | 19 |
| v1 | 0.5964 | 21 | 0.1161 | 21 |

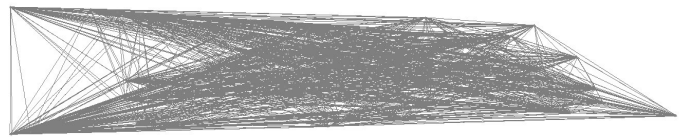

Fig. 5. A 50 node graph used for comparison of the centrality measures.

TABLE 3
COMPARISON BETWEEN EIGENVECTOR CENTRALITY AND PAGERANK CENTRALITY FOR THE GRAPH SHOWN IN FIG. 5

| Vertex | PageRank | Rank | Eigenvector | Rank |
|---|---|---|---|---|
| v1 | 1.2113 | 1 | 0.178 | 1 |
| v28 | 1.2057 | 2 | 0.1746 | 2 |
| v11 | 1.1892 | 3 | 0.1733 | 3 |

| Vertex | Page rank | Rank | Eigenvector | Rank |
|---|---|---|---|---|
| v48 | 1.1889 | 4 | 0.1715 | 4 |
| v42 | 1.1646 | 5 | 0.1652 | 5 |
| v8 | 1.1379 | 6 | 0.162 | 6 |
| v5 | 1.1314 | 7 | 0.1609 | 7 |
| v4 | 1.1222 | 8 | 0.1596 | 9 |
| v21 | 1.1181 | 9 | 0.1606 | 8 |
| v2 | 1.1096 | 10 | 0.1561 | 11 |
| v50 | 1.1041 | 11 | 0.1587 | 10 |
| v25 | 1.0955 | 12 | 0.1558 | 12 |
| v34 | 1.0772 | 13 | 0.1517 | 14 |
| v39 | 1.0678 | 14 | 0.1497 | 17 |
| v44 | 1.065 | 15 | 0.1531 | 13 |
| v27 | 1.064 | 16 | 0.1496 | 18 |
| v7 | 1.0609 | 17 | 0.1434 | 24 |
| v30 | 1.0606 | 18 | 0.15 | 15 |
| v22 | 1.0486 | 19 | 0.1498 | 16 |
| v35 | 1.0408 | 20 | 0.1457 | 20 |
| v41 | 1.0335 | 21 | 0.1458 | 19 |
| v9 | 1.0331 | 22 | 0.144 | 23 |
| v38 | 1.0297 | 23 | 0.1446 | 22 |
| v3 | 1.0248 | 24 | 0.1422 | 26 |
| v36 | 1.0225 | 25 | 0.1452 | 21 |
| v16 | 1.0168 | 26 | 0.1404 | 28 |
| v6 | 1.015 | 27 | 0.1431 | 25 |
| v24 | 1.0089 | 28 | 0.1399 | 30 |
| v13 | 1.0043 | 29 | 0.1418 | 27 |
| v17 | 0.9951 | 30 | 0.14 | 29 |
| v32 | 0.9915 | 31 | 0.1399 | 30 |
| v37 | 0.986 | 32 | 0.1393 | 32 |
| v20 | 0.9787 | 33 | 0.1343 | 33 |
| v47 | 0.9483 | 34 | 0.1295 | 36 |
| v10 | 0.9298 | 35 | 0.1319 | 34 |
| v23 | 0.929 | 36 | 0.129 | 37 |
| v45 | 0.9248 | 37 | 0.131 | 35 |
| v31 | 0.9098 | 38 | 0.1279 | 38 |
| v26 | 0.8998 | 39 | 0.1249 | 39 |
| v33 | 0.8924 | 40 | 0.1227 | 40 |
| v15 | 0.8881 | 41 | 0.1214 | 41 |
| v49 | 0.8583 | 42 | 0.1151 | 42 |
| Vertex | Page rank | Rank | Eigenvector | Rank |
| v19 | 0.8319 | 43 | 0.1151 | 42 |
| v12 | 0.8243 | 44 | 0.1144 | 44 |
| v46 | 0.8193 | 45 | 0.1102 | 45 |
| v18 | 0.8165 | 46 | 0.1047 | 48 |
| v43 | 0.8086 | 47 | 0.1046 | 49 |
| v40 | 0.7931 | 48 | 0.1074 | 46 |
| v14 | 0.7762 | 49 | 0.1051 | 47 |
| v29 | 0.7466 | 50 | 0.0961 | 50 |

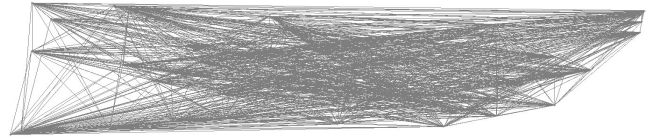

Fig. 6. A 50 node graph used for comparison of the centrality measures.

TABLE 4
COMPARISON BETWEEN EIGENVECTOR CENTRALITY AND PAGERANK CENTRALITY FOR THE GRAPH SHOWN IN FIG. 6

| Vertex | PageRank | Rank | Eigenvector | Rank |
|---|---|---|---|---|
| v30 | 1.2687 | 1 | 0.1821 | 1 |
| v12 | 1.2547 | 2 | 0.1811 | 3 |
| v11 | 1.2358 | 3 | 0.1814 | 2 |
| v5 | 1.1997 | 4 | 0.1739 | 4 |
| v50 | 1.1722 | 5 | 0.1654 | 5 |
| v2 | 1.1581 | 6 | 0.1641 | 7 |
| v6 | 1.1392 | 7 | 0.1651 | 6 |
| v32 | 1.1296 | 8 | 0.1623 | 8 |
| v8 | 1.1268 | 9 | 0.1614 | 9 |
| v27 | 1.1218 | 10 | 0.1585 | 10 |
| v24 | 1.1084 | 11 | 0.1568 | 11 |
| v43 | 1.0988 | 12 | 0.1523 | 16 |
| v9 | 1.0899 | 13 | 0.1568 | 11 |
| v4 | 1.0839 | 14 | 0.1548 | 14 |
| v35 | 1.0799 | 15 | 0.1505 | 17 |
| v23 | 1.0786 | 16 | 0.1526 | 15 |
| v45 | 1.0778 | 17 | 0.1551 | 13 |
| v22 | 1.0636 | 18 | 0.1486 | 19 |
| v28 | 1.0577 | 19 | 0.1466 | 21 |
| v20 | 1.0549 | 20 | 0.1501 | 18 |
| v31 | 1.044 | 21 | 0.148 | 20 |
| v44 | 1.031 | 22 | 0.1447 | 22 |
| v41 | 1.0235 | 23 | 0.1417 | 23 |
| v36 | 1.0091 | 24 | 0.1371 | 26 |
| v15 | 1.0058 | 25 | 0.1414 | 24 |

| Vertex | Page rank | Rank | Eigenvector | Rank |
|---|---|---|---|---|
| v33 | 1.002 | 26 | 0.1384 | 25 |
| v21 | 0.9746 | 27 | 0.1367 | 27 |
| v26 | 0.966 | 28 | 0.1331 | 28 |
| v3 | 0.9568 | 29 | 0.131 | 31 |
| v39 | 0.9422 | 30 | 0.1325 | 30 |
| v10 | 0.9409 | 31 | 0.1305 | 32 |
| v42 | 0.9397 | 32 | 0.1327 | 29 |
| v25 | 0.9256 | 33 | 0.1291 | 33 |
| v47 | 0.9227 | 34 | 0.127 | 35 |
| v34 | 0.9205 | 35 | 0.1275 | 34 |
| v13 | 0.9203 | 36 | 0.1265 | 36 |
| v48 | 0.8949 | 37 | 0.1215 | 37 |
| v1 | 0.8937 | 38 | 0.1179 | 42 |
| v29 | 0.8916 | 39 | 0.1196 | 40 |
| Vertex | Page rank | Rank | Eigenvector | Rank |
| v37 | 0.8728 | 40 | 0.1203 | 39 |
| v16 | 0.8691 | 41 | 0.1166 | 46 |
| v18 | 0.8634 | 42 | 0.1173 | 43 |
| v7 | 0.8631 | 43 | 0.1171 | 44 |
| v38 | 0.8534 | 44 | 0.1215 | 37 |
| v14 | 0.8439 | 45 | 0.1171 | 44 |
| v40 | 0.8405 | 46 | 0.1191 | 41 |
| v19 | 0.8368 | 47 | 0.1148 | 48 |
| v46 | 0.8245 | 48 | 0.1151 | 47 |
| v49 | 0.8202 | 49 | 0.1103 | 49 |
| v17 | 0.7071 | 50 | 0.0896 | 50 |

TABLE 5
COMPARISON BETWEEN EIGENVECTOR CENTRALITY AND PAGERANK CENTRALITY FOR A 100 NODE GRAPH

| Vertex | PageRank | Rank | Eigenvector | Rank |
|---|---|---|---|---|
| v8 | 1.254 | 1 | 0.1294 | 1 |
| v24 | 1.1787 | 2 | 0.1196 | 2 |
| v55 | 1.1681 | 3 | 0.1185 | 3 |
| v28 | 1.1599 | 4 | 0.1165 | 4 |
| v61 | 1.1446 | 5 | 0.1164 | 5 |
| v62 | 1.1398 | 6 | 0.1148 | 7 |
| v91 | 1.1364 | 7 | 0.1138 | 8 |
| v1 | 1.1207 | 8 | 0.1133 | 9 |
| v6 | 1.117 | 9 | 0.1151 | 6 |
| v76 | 1.1141 | 10 | 0.113 | 10 |
| v48 | 1.112 | 11 | 0.1112 | 12 |
| v52 | 1.1044 | 12 | 0.1099 | 16 |
| v9 | 1.1004 | 13 | 0.1117 | 11 |
| v85 | 1.0982 | 14 | 0.1098 | 17 |
| v56 | 1.0928 | 15 | 0.1101 | 14 |
| v94 | 1.0899 | 16 | 0.1106 | 13 |
| v100 | 1.0845 | 17 | 0.1101 | 14 |
| v2 | 1.082 | 18 | 0.1088 | 19 |
| v83 | 1.0805 | 19 | 0.1096 | 18 |
| v84 | 1.0691 | 20 | 0.1073 | 22 |
| v87 | 1.0672 | 21 | 0.1088 | 19 |
| v26 | 1.0669 | 22 | 0.1069 | 24 |
| v34 | 1.0667 | 23 | 0.1064 | 25 |
| v11 | 1.0658 | 24 | 0.1074 | 21 |
| v4 | 1.0642 | 25 | 0.1057 | 28 |
| v95 | 1.0578 | 26 | 0.107 | 23 |
| v53 | 1.0558 | 27 | 0.1056 | 29 |
| v68 | 1.0544 | 28 | 0.1047 | 33 |
| v71 | 1.054 | 29 | 0.1063 | 26 |
| v75 | 1.0536 | 30 | 0.1051 | 30 |
| v81 | 1.0498 | 31 | 0.105 | 32 |
| v86 | 1.0479 | 32 | 0.104 | 38 |
| v35 | 1.0473 | 33 | 0.1045 | 35 |
| v12 | 1.0454 | 34 | 0.1059 | 27 |
| v27 | 1.0452 | 35 | 0.1047 | 33 |
| v54 | 1.0442 | 36 | 0.1051 | 30 |
| v21 | 1.041 | 37 | 0.1043 | 36 |
| v89 | 1.0405 | 38 | 0.1021 | 42 |
| v15 | 1.0393 | 39 | 0.104 | 38 |
| v80 | 1.0309 | 40 | 0.1043 | 36 |
| v30 | 1.0294 | 41 | 0.102 | 43 |
| v22 | 1.0274 | 42 | 0.1023 | 41 |
| Vertex | PageRank | Rank | Eigenvector | Rank |
| v57 | 1.0176 | 44 | 0.1006 | 45 |
| v82 | 1.0165 | 45 | 0.1024 | 40 |
| v44 | 1.0159 | 46 | 0.1006 | 45 |
| v25 | 1.0137 | 47 | 0.1004 | 47 |
| v77 | 1.0088 | 48 | 0.1004 | 47 |
| v41 | 1.0007 | 49 | 0.099 | 52 |
| v32 | 0.9993 | 50 | 0.0983 | 57 |
| v5 | 0.9989 | 51 | 0.1004 | 47 |
| v78 | 0.9925 | 52 | 0.0986 | 55 |

| Vertex | | | | | Vertex | | | | |
|---|---|---|---|---|---|---|---|---|---|
| v98 | 0.9912 | 53 | 0.0987 | 53 | v46 | 0.855 | 94 | 0.085 | 89 |
| v14 | 0.9908 | 54 | 0.0972 | 60 | v38 | 0.8536 | 95 | 0.0838 | 93 |
| v66 | 0.9906 | 55 | 0.0986 | 55 | v69 | 0.8368 | 96 | 0.0796 | 97 |
| v63 | 0.9905 | 56 | 0.098 | 58 | v40 | 0.8301 | 97 | 0.0805 | 96 |
| v39 | 0.9879 | 57 | 0.0994 | 50 | v17 | 0.8224 | 98 | 0.0789 | 98 |
| v20 | 0.9878 | 58 | 0.0991 | 51 | v23 | 0.8188 | 99 | 0.0781 | 99 |
| v74 | 0.9827 | 59 | 0.0973 | 59 | v96 | 0.7894 | 100 | 0.0743 | 100 |
| v92 | 0.9825 | 60 | 0.0971 | 62 | | | | | |
| v58 | 0.9816 | 61 | 0.0987 | 53 | | | | | |
| v33 | 0.9789 | 62 | 0.096 | 65 | | | | | |
| v59 | 0.9778 | 63 | 0.0967 | 63 | | | | | |
| v64 | 0.9744 | 64 | 0.0972 | 60 | | | | | |
| v73 | 0.9707 | 65 | 0.0967 | 63 | | | | | |
| v97 | 0.9676 | 66 | 0.0955 | 67 | | | | | |
| v36 | 0.9601 | 67 | 0.0958 | 66 | | | | | |
| v37 | 0.9562 | 68 | 0.0934 | 73 | | | | | |
| v88 | 0.9562 | 68 | 0.0947 | 69 | | | | | |
| v51 | 0.955 | 70 | 0.0941 | 72 | | | | | |
| v19 | 0.9548 | 71 | 0.0953 | 68 | | | | | |
| v60 | 0.9475 | 72 | 0.0942 | 71 | | | | | |
| v45 | 0.9474 | 73 | 0.0943 | 70 | | | | | |
| v43 | 0.9463 | 74 | 0.0928 | 74 | | | | | |
| v50 | 0.9455 | 75 | 0.0917 | 76 | | | | | |
| v42 | 0.9451 | 76 | 0.0918 | 75 | | | | | |
| v47 | 0.9403 | 77 | 0.0915 | 77 | | | | | |
| v18 | 0.9269 | 78 | 0.0901 | 83 | | | | | |
| v3 | 0.9234 | 79 | 0.0915 | 77 | | | | | |
| v72 | 0.9232 | 80 | 0.0909 | 79 | | | | | |
| v90 | 0.9186 | 81 | 0.0907 | 80 | | | | | |
| v13 | 0.9176 | 82 | 0.0907 | 80 | | | | | |
| v67 | 0.9171 | 83 | 0.0901 | 83 | | | | | |
| v29 | 0.9012 | 84 | 0.0881 | 86 | | | | | |
| v31 | 0.9012 | 84 | 0.0903 | 82 | | | | | |
| v99 | 0.8988 | 86 | 0.0873 | 87 | | | | | |
| v65 | 0.8973 | 87 | 0.0897 | 85 | | | | | |
| v79 | 0.8888 | 88 | 0.0872 | 88 | | | | | |
| v70 | 0.8824 | 89 | 0.0847 | 91 | | | | | |
| v7 | 0.8774 | 90 | 0.0849 | 90 | | | | | |
| v16 | 0.8729 | 91 | 0.084 | 92 | | | | | |
| v10 | 0.8585 | 92 | 0.0833 | 95 | | | | | |
| v49 | 0.856 | 93 | 0.0835 | 94 | | | | | |

TABLE 6
COMPARISON BETWEEN EIGENVECTOR CENTRALITY AND PAGERANK CENTRALITY FOR A 100 NODE GRAPH

| Vertex | PageRank | Rank | Eigenvector | Rank |
|---|---|---|---|---|
| v51 | 1.1992 | 1 | 0.1222 | 1 |
| v77 | 1.1976 | 2 | 0.1215 | 2 |
| v47 | 1.1793 | 3 | 0.1203 | 3 |
| v50 | 1.1453 | 4 | 0.1167 | 5 |
| v11 | 1.1409 | 5 | 0.1169 | 4 |
| v42 | 1.1263 | 6 | 0.114 | 7 |
| v82 | 1.1257 | 7 | 0.1131 | 9 |
| v73 | 1.1236 | 8 | 0.1141 | 6 |
| v66 | 1.1204 | 9 | 0.1132 | 8 |
| v17 | 1.1189 | 10 | 0.1123 | 10 |
| v100 | 1.1024 | 11 | 0.111 | 11 |
| v21 | 1.0941 | 12 | 0.1099 | 14 |
| v7 | 1.0884 | 13 | 0.1107 | 12 |
| v31 | 1.0882 | 14 | 0.1099 | 14 |
| v41 | 1.0827 | 15 | 0.1075 | 23 |
| v84 | 1.0814 | 16 | 0.1071 | 24 |
| v98 | 1.0806 | 17 | 0.1099 | 14 |
| v86 | 1.0791 | 18 | 0.11 | 13 |
| v90 | 1.0782 | 19 | 0.1086 | 18 |
| v27 | 1.0774 | 20 | 0.1093 | 17 |
| v70 | 1.0744 | 21 | 0.1077 | 21 |
| v69 | 1.0682 | 22 | 0.1077 | 21 |
| v32 | 1.0671 | 23 | 0.1081 | 19 |
| v38 | 1.0651 | 24 | 0.1078 | 20 |
| v97 | 1.0603 | 25 | 0.1043 | 34 |
| v58 | 1.0594 | 26 | 0.1066 | 25 |
| v62 | 1.0562 | 27 | 0.1056 | 30 |
| v78 | 1.0538 | 28 | 0.106 | 28 |
| v52 | 1.0522 | 29 | 0.1063 | 26 |
| v80 | 1.0522 | 29 | 0.1062 | 27 |

| Vertex | PageRank | Rank | Eigenvector | Rank |
|---|---|---|---|---|
| v44 | 1.0517 | 31 | 0.104 | 37 |
| v28 | 1.0476 | 32 | 0.1059 | 29 |
| v36 | 1.0414 | 33 | 0.1047 | 31 |
| v9 | 1.0359 | 34 | 0.1039 | 39 |
| v63 | 1.0335 | 35 | 0.1038 | 40 |
| v89 | 1.0329 | 36 | 0.1043 | 34 |
| v92 | 1.0329 | 36 | 0.1044 | 32 |
| v30 | 1.0305 | 38 | 0.1044 | 32 |
| v19 | 1.029 | 39 | 0.104 | 37 |
| v43 | 1.0273 | 40 | 0.1024 | 43 |
| v94 | 1.0265 | 41 | 0.1041 | 36 |
| v85 | 1.0258 | 42 | 0.1019 | 45 |
| v35 | 1.0252 | 43 | 0.1025 | 42 |
| v79 | 1.0247 | 44 | 0.1036 | 41 |
| v49 | 1.0233 | 45 | 0.1017 | 46 |
| v60 | 1.0195 | 46 | 0.1006 | 49 |
| v55 | 1.018 | 47 | 0.1014 | 47 |
| v6 | 1.0165 | 48 | 0.1 | 52 |
| v24 | 1.0139 | 49 | 0.1014 | 47 |
| v54 | 1.0078 | 50 | 0.1021 | 44 |
| v53 | 1.0024 | 51 | 0.1004 | 50 |
| v1 | 1.0016 | 52 | 0.1001 | 51 |
| v61 | 1.0013 | 53 | 0.0994 | 54 |
| v16 | 0.9985 | 54 | 0.1 | 52 |
| v46 | 0.9953 | 55 | 0.0988 | 56 |
| v25 | 0.9916 | 56 | 0.0991 | 55 |
| Vertex | PageRank | Rank | Eigenvector | Rank |
| v5 | 0.9872 | 58 | 0.0973 | 59 |
| v40 | 0.9787 | 59 | 0.0987 | 57 |
| v3 | 0.9736 | 60 | 0.0959 | 60 |
| v4 | 0.9728 | 61 | 0.0954 | 65 |
| v18 | 0.9696 | 62 | 0.0956 | 63 |
| v95 | 0.9666 | 63 | 0.0958 | 62 |
| v2 | 0.966 | 64 | 0.0951 | 67 |
| v59 | 0.9649 | 65 | 0.0959 | 60 |
| v14 | 0.9619 | 66 | 0.0943 | 71 |
| v13 | 0.9613 | 67 | 0.0946 | 68 |
| v67 | 0.9611 | 68 | 0.0936 | 73 |
| v75 | 0.9599 | 69 | 0.0955 | 64 |
| v68 | 0.9597 | 70 | 0.0954 | 65 |
| v15 | 0.959 | 71 | 0.0946 | 68 |
| v22 | 0.9555 | 72 | 0.0928 | 76 |
| v23 | 0.9541 | 73 | 0.0935 | 74 |
| v88 | 0.9504 | 74 | 0.0933 | 75 |
| v34 | 0.9502 | 75 | 0.0924 | 77 |
| v91 | 0.9471 | 76 | 0.0945 | 70 |
| v83 | 0.9451 | 77 | 0.0939 | 72 |
| v20 | 0.9413 | 78 | 0.0924 | 77 |
| v96 | 0.9389 | 79 | 0.0916 | 80 |
| v93 | 0.9309 | 80 | 0.0913 | 81 |
| v26 | 0.9248 | 81 | 0.092 | 79 |
| v10 | 0.9207 | 82 | 0.0892 | 83 |
| v99 | 0.9188 | 83 | 0.0885 | 84 |
| v76 | 0.9117 | 84 | 0.0893 | 82 |
| v45 | 0.9083 | 85 | 0.0884 | 85 |
| v37 | 0.9062 | 86 | 0.0872 | 89 |
| v56 | 0.9019 | 87 | 0.0876 | 87 |
| v74 | 0.8942 | 88 | 0.0874 | 88 |
| v72 | 0.8925 | 89 | 0.0877 | 86 |
| v12 | 0.8825 | 90 | 0.0856 | 91 |
| v8 | 0.8806 | 91 | 0.0863 | 90 |
| v33 | 0.8731 | 92 | 0.0845 | 92 |
| v87 | 0.8683 | 93 | 0.0835 | 93 |
| v64 | 0.8615 | 94 | 0.0832 | 95 |
| v71 | 0.851 | 95 | 0.0816 | 96 |
| v81 | 0.8507 | 96 | 0.0833 | 94 |
| v65 | 0.8339 | 97 | 0.0801 | 97 |
| v39 | 0.8208 | 98 | 0.0793 | 98 |
| v29 | 0.8084 | 99 | 0.0775 | 99 |
| v57 | 0.7512 | 100 | 0.0708 | 100 |

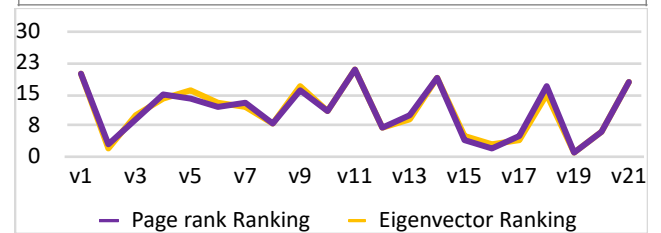
Fig. 7. A graph showing comparison of the ranks for Fig. 3.

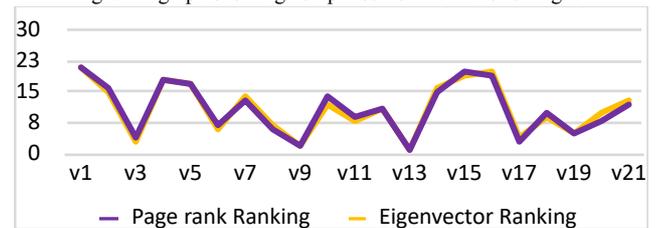
Fig. 8. A graph showing comparison of the ranks for Fig. 4.

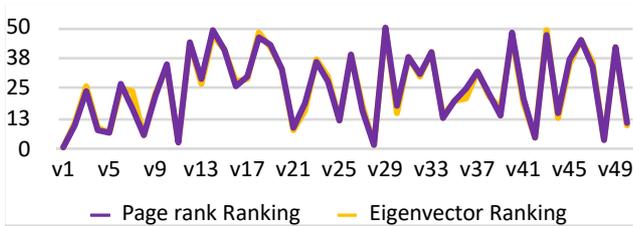
Fig. 9. A graph showing comparison of the ranks for Fig. 5.

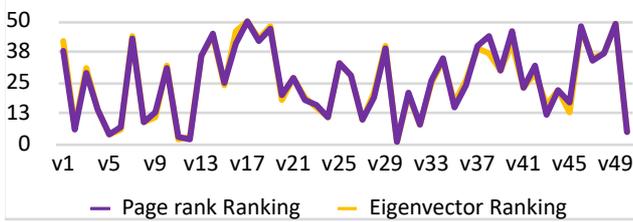
Fig. 10. A graph showing comparison of the ranks for Fig. 6

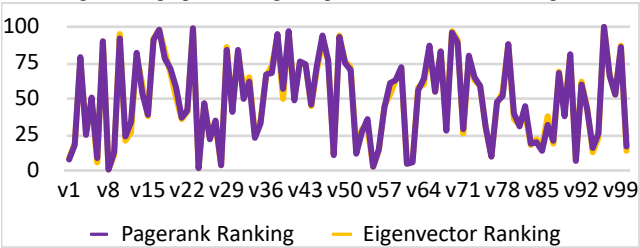
Fig. 11. A graph showing comparison of the ranks for Table 5.

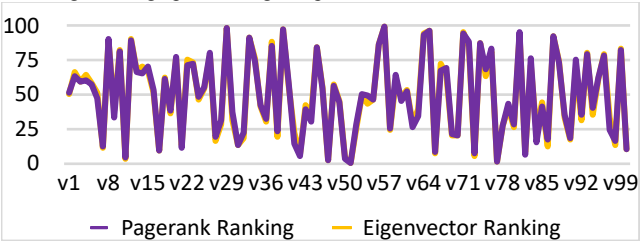
Fig. 12. A graph showing comparison of the ranks for Table 6

From Table 1-Table 6, showing the comparison of the ranks assigned by application of the eigenvector formula is comparable to the ranks obtained by PageRank. In a search engine that will be used for finding the most accurate results, the highest priority is given to the first few pages as those are the ones that will attract the attention of the user who is surfing the internet. For all the graphs taken, the first five vertices analogous to the webpages ranking are the same, thus strengthening our argument for the use of Eigenvector in place of PageRank. The figures (Fig. 7- Fig. 12) show that the variation of the ranks assigned by eigenvector centrality as compared to the ones assigned using the PageRank algorithm. As observed in the figures, the variation is insignificant and it goes on decreasing with the increase in the number of nodes in the graph. Therefore, in an actual web graph the difference would be reduced to a miniscule amount while keeping the top nodes or the top pages the same.

## V. SPEARMAN'S RANK COEFFICIENT CORRELATION

The Spearman's rank correlation coefficient is a nonparametric measure that determines the statistical dependence of one variable on the other. It uses a monotonic function to assess the relation between the two variables. It measures how strongly the two variables which are ranked are associated.

## VI. PEARSON'S COEFFICIENT CORRELATION

The Pearson's rank correlation coefficient is much like Spearman's coefficient correlation however it differs in the respect that it indicates the linear relation between the two variables.

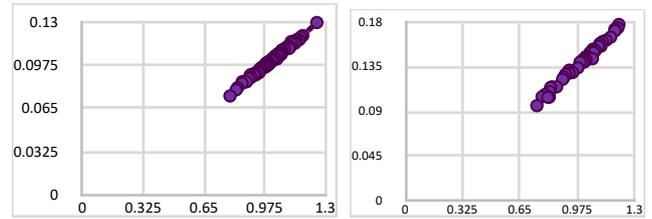
Fig. 13. Graphs showing comparison of PageRank and Eigenvector

As shown in the figures Fig. 13, the graph between PageRank and Eigenvector centrality is not only monotonic, but is also linear. Hence, it gives a value of almost 1 for both the Pearson's coefficient correlation test and Spearman's coefficient correlation test. Some of the values obtained by Spearman's coefficient correlation to back the hypothesis that PageRank Centrality can be easily replaced with eigenvector centrality are given in Table 7. Critical value for 21 node graph is 0.681 for a 0.05% probability that the value occurred by chance. Similarly for 50 nodes graph, the critical value is 0.465 for a 0.05% probability that the value occurred by chance and for a 100 node graph, the critical value is 0.326 for a 0.05% probability that the value occurred by chance.

TABLE 7
SPEARMAN'S RANK COEFFICIENT CORRELATION VALUES

| Number Of nodes | Spearman's Rank coefficient correlation | Number Of nodes | Spearman's Rank coefficient correlation |
|---|---|---|---|
| 21 | 0.979220779 | 50 | 0.9833613 |
| 21 | 0.928571429 | 50 | 0.9822449 |
| 21 | 0.940259740 | 50 | 0.9923649 |
| 21 | 0.941558442 | 50 | 0.9884514 |
| 21 | 0.988311688 | 50 | 0.9909484 |
| 21 | 0.987012987 | 50 | 0.9833613 |
| 21 | 0.954545455 | 50 | 0.9884994 |
| 21 | 0.979220779 | 50 | 0.9826170 |
| 21 | 0.985714286 | 50 | 0.9791357 |
| 21 | 0.945454545 | 100 | 0.9951995 |

| Number Of nodes | Spearman's Rank coefficient correlation | Number Of nodes | Spearman's Rank coefficient correlation |
|---|---|---|---|
| 50 | 0.989531813 | 100 | 0.9674767 |
| 50 | 0.992364946 | 100 | 0.9945875 |
| 50 | 0.990948379 | 100 | 0.9945875 |

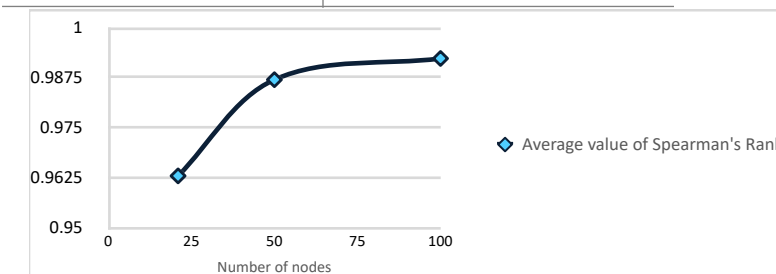

Fig. 15. A scatterplot representing the average value of Spearman's Rank Coefficient Correlation for different number of nodes.

As shown in Fig. 15, the average value of Spearman's Rank Coefficient correlation increases with the increase in the number of nodes in the graph. This shows that the strength of association between the values obtained by eigenvector and PageRank continue to converge with the increase in the number of nodes, analogous to webpages, which is practically $n \to \infty$.

## VII. CONCLUSION

In a fast paced world where each nanosecond has proved to be of crucial importance, it is essential to adopt every possible means to save time. With this purpose in mind, we have proved the dominance of eigenvector centrality over the existing PageRank algorithm in the respect of time complexity showing positive results in our favour by giving supportive and conclusive evidence.

## VIII. REFERENCES

1. Luca Donetti, Franco Neri and Miguel A Muoz "Optimal network topologies:expanders, cages, Ramanujan graphs, entangled networks and all that" Journal of Statistical Mathematics (2006)
2. Sepandar D. Kamvar. Mario T. Schlosser, Hector Garcia-Molina, "The EigenTrust Algorithm for Reputation Management in P2P Networks," Stanford University
3. Z.Gyöngyi, H. Garcia-Molina, J.Pedersen "Combating Web Spam with Trust Rank" Stanford University, thirtieth international conference on very large databases – volume 30
4. Krishnan, Vijay; Raj, Rashmi. "Web Spam Detection with Anti-Trust Rank" Stanford University.
5. S.Brin and L.Page, "The PageRank Citation Ranking: Bringing Order to the Web" Stanford InfoLab, (Jan.,1998)
6. Gabor Csardi, Tamas Nepusz, "The igraph software package for complex network research," InterJournal Complex Systems 2006
7. Spizzirri, Leo, 2011, Justification and application of eigenvector centrality, working paper.
8. Gabriele Lohmann , Daniel S. Margulies, Annette Horstmann, Burkhard Pleger, Joeran Lepsien, Dirk Goldhahn, Haiko Schloegl, Michael Stumvoll, Arno Villringer, Robert Turner "Eigenvector Centrality Mapping for Analyzing Connectivity Patterns in fMRI Data of the Human Brain"
9. B.Jaganathan,Kalyani Desikan, "Category-Based Pagerank Algorithm," International Journal of Pure and Applied Mathematics., vol.101,No.5, pp. 811-820, August 2015.
10. B.Jaganathan,Kalyani Desikan, "Penalty-Based Pagerank Algorithm," ARPN Journal of Engineering and Applied Sciences, vol.10,No.5
11. B.Jaganathan,Kalyani Desikan, "Weighted Pagerank Algorithm based on In-Out weight of webpages," Indian Journal of Science and Technology, vol.8,No.34, pp. 1-6, December 2015